\documentclass{article}
\usepackage{amssymb,amsmath}
\usepackage{epsfig}
\newtheorem{rem}{Remark}[section]
\newcommand{\br}{\begin{rem}}
\newcommand{\er}{\end{rem}}
\newtheorem{ex}{Example}[section]
\newcommand{\bex}{\begin{ex}}
\newcommand{\eex}{\end{ex}}
\newtheorem{Def}{Definition}[section]
\newcommand{\bd}{\begin{Def}}
\newcommand{\ed}{\end{Def}}
\newtheorem{theorem}{Theorem}[section]
\newcommand{\bt}{\begin{theorem}}
\newcommand{\et}{\end{theorem}}
\newtheorem{lemma}{Lemma}[section]
\newcommand{\bl}{\begin{lemma}}
\newcommand{\el}{\end{lemma}}
\newcommand{\be}{\begin{equation}}
\newcommand{\ee}{\end{equation}}
\newcommand{\bea}{\begin{eqnarray}}
\newcommand{\eea}{\end{eqnarray}}
\newcommand{\pa}{\partial}
\newcommand{\nn}{\nonumber}
\newcommand{\adots}{\mathinner{\mkern2mu\raise1pt\hbox{.}\mkern2mu
\raise4pt\hbox{.}\mkern2mu\raise7pt\hbox{.}\mkern1mu}}

\title{A note on some superintegrable Hamiltonian systems}
\author{Allan Fordy, School of Mathematics, \\
University of Leeds\\
e-mail a.p.fordy@leeds.ac.uk}%\date{}

\begin{document}

\maketitle

\begin{abstract}
We consider some examples of superintegrable system which were recently isolated through a differential Galois group analysis.
The identity of these systems is clarified and the corresponding Poisson algebras derived.
\end{abstract}



\section{Introduction}

A Hamiltonian system of $n$ degrees of freedom, Hamiltonian $H$, is said to be {\em completely integrable in the Liouville sense} if we have $n$ independent functions $I_n$, which are {\em in involution} (mutually Poisson commuting), with $H$ being a function of these and typically just one of them. Whilst $n$ is the maximal number of independent functions which can be {\em in involution}, it is possible to have further integrals of the Hamiltonian $H$, which necessarily generate a non-Abelian algebra of integrals of $H$.  The maximal number of additional {\em independent} integrals is $n-1$, since the ``level surface'' of $2n-1$ integrals (meaning the intersection of individual level surfaces) is just the (unparameterised) integral curve.  Well known elementary examples are the isotropic harmonic oscillator, the Kepler system and the Calogero-Moser system.  The quadratures of complete integrability are often achieved through the separation of variables of the Hamilton-Jacobi equation.  The solution of a maximally super-integrable system can also be calculated purely algebraically (albeit implicitly), requiring just the solution of the equations $I_k=c_k,\, k=1,\dots ,2n-1$.  Maximally superintegrable systems have a number of interesting properties: they can be separable in more than one coordinate system; all bounded orbits are closed; they give rise to interesting Poisson algebras with polynomial Poisson relations.  The idea can be extended to {\em quantum integrable systems}, with first integrals replaced by commuting differential operators. For some examples of superintegrable quantum systems it is possible to use the additional commuting operators to build sequences of eigenfunctions \cite{f07-1,f13-1}.  There is a large literature on the classification and analysis of superintegrable systems (see the review \cite{13-3}) and they naturally occur in many applications in physics (additional integrals being referred to as ``hidden symmetries'') \cite{14-2}.

A superintegrable system may occur in some other context and not be written in any recognisable form, so requires some judicious change of coordinates.  In \cite{15-1} the authors considered Hamiltonian functions of the form
\be\label{rmk-ham}
H=\frac{1}{2} r^{m-k} \left(p_r^2+\frac{p_\varphi^2}{r^2}\right) + r^m U(\varphi),
\ee
using differential Galois theory to determine necessary conditions for complete integrability.
Restricting attention to the case $U(\varphi)=-\cos\varphi$, they found $8$ cases, four of which were not only completely integrable, but also \underline{superintegrable} (see their list (3.8)).  In fact, for each case, they also presented $2$ first integrals, $F_1$ and $F_2$, which are \underline{quadratic} in momenta, which means that each of these systems separates in $2$ different coordinate systems.

In this paper we consider these $4$ examples in more detail.  The motivation of \cite{15-1} was to extend the ideas of ``homogeneous systems'' to curved spaces, although the metric implied by the kinetic energy of (\ref{rmk-ham}) is, in fact, flat.  A transformation to flat coordinates is derived in Section \ref{sect-flat}, giving us $3$ ``geometric'' first integrals of the kinetic energy (related to the Killing vectors of the Euclidean metric), which are used to build quadratic integrals (specifically chosen to reproduce the examples of \cite{15-1}).  In fact, they share a common first integral, which allows us to use parabolic coordinates as  common separation variables.  The existence of the additional first integral in the superintegrable cases restricts the general separable potential to a finite parameter family.  The $4$ examples belong to two such superintegrable reductions, discussed in Section \ref{sect-sep}.  It is shown that $2$ of the examples actually have a first order integral.  The results of Section \ref{sect-sep} are translated back into the original coordinates in Section \ref{sect-polar} and compared with \cite{15-1}.

\section{Flat Coordinates for the Metric of (\ref{rmk-ham})}\label{sect-flat}

It is easy to check that the inverse of the kinetic energy tensor of (\ref{rmk-ham}) defines a flat metric.  Rather than compute the curvature, we calculate flat (Euclidean) coordinates $(Q_1,Q_2)$.  We first summarise the algebraic properties of the Euclidean algebra (the algebra of Killing vectors of the Euclidean metric), working within the Hamiltonian framework.

\subsection{The Euclidean Algebra}

In a Euclidean space, with coordinates $(Q_1,Q_2)$ and conjugate momenta $(P_1,P_2)$, the kinetic energy takes the form
\be\label{te}
T_e = \frac{1}{2}\, (P_1^2+P_2^2).
\ee
The coordinates $Q_i$ satisfy the simple relations
\be\label{qhh}
\{\{Q_i,T_e\},T_e\}=0,\quad P_i=\{Q_i,T_e\},\quad \{Q_i,\{Q_i,T_e\}\}=1.
\ee
The Hamiltonian $T_e$ has $3$ first degree integrals
\begin{subequations}\label{Euclid}
\be\label{k123}
K_1=P_1,\quad K_2=P_2,\quad K_3=Q_2P_1-Q_1P_2,
\ee
which satisfy the usual Euclidean algebra Poisson relations:
\be\label{k1k2}
\{K_1,K_2\} = 0, \quad \{K_1,K_3\} = K_2, \quad \{K_2,K_3\} = -K_1,
\ee
\end{subequations}
with Casimir $T_e=\frac{1}{2} (K_1^2+K_2^2)$, and correspond to the $3$ Killing vectors of the $2D$ Euclidean metric:
$$
v_i = \{Q_1,K_i\} \frac{\pa}{\pa Q_1} + \{Q_2,K_i\} \frac{\pa}{\pa Q_2}\quad\Rightarrow\quad   [v_i,v_j]=-v_{\small \{K_i,K_j\}}.
$$

\subsection{Transformation between $(r,\varphi)$ and $(Q_1,Q_2)$}

Written in terms of any other coordinate system, $K_i$ must satisfy the same relations.  Starting with
\be\label{tr}
T_r = \frac{1}{2}\, r^n \left(p_r^2+\frac{p_\varphi^2}{r^2}\right),
\ee
we may consider $\{\{Q,T_r\},T_r\}=0$ as a system of {\em partial differential equations} for function $Q(r,\varphi)$ (the coefficients of $p_r^2,\, p_r p_\varphi,\, p_\varphi^2$):
$$
\left(r^{\frac{n}{2}} Q\right)_r=0,\quad \left(r^{\frac{n-2}{2}} Q\right)_{r\varphi}=0,\quad 2 Q_{\varphi\varphi}-(n-2) r Q_r=0.
$$
This is an overdetermined system whose integrability conditions are guaranteed by \underline{flatness}.  The solution contains two essential parameters, so we define
\begin{subequations}\label{Qr-tran}
\be\label{Qr}
Q_1= b_1 r^{1-n/2} \cos \frac{(n-2)\varphi}{2},\quad
    Q_2= b_2 r^{1-n/2} \sin \frac{(n-2)\varphi}{2},\quad n\neq 2.
\ee
The relation $\{Q_i,\{Q_i,T_r\}\}=1$ implies $b_1=b_2=2/(2-n)$, and
\bea
P_1 &=& \{Q_1,T_r\} = r^{n/2} \left(p_r\, \cos \frac{(n-2)\varphi}{2}+
                   \frac{p_\varphi}{r}\, \sin \frac{(n-2)\varphi}{2}\right),\nn\\
                   \label{p1p2}  \\
P_2 &=& \{Q_2,T_r\} =r^{n/2} \left(p_r\, \sin \frac{(n-2)\varphi}{2}-
                   \frac{p_\varphi}{r}\, \cos \frac{(n-2)\varphi}{2}\right).\nn
\eea
\end{subequations}
In these coordinates, $K_1=P_1$ and $K_2=P_2$ (now defined by (\ref{p1p2})) and $K_3=\frac{2 p_\varphi}{(2-n)}$, which, of course, satisfy the same Poisson relations (\ref{k1k2}), with Casimir $T_r=\frac{1}{2} (K_1^2+K_2^2)$.

\br
There is an alternative solution for the case $n=2$, but this is not of interest to us here.
\er

\section{Separability and Superintegrability}\label{sect-sep}

We start with the standard total energy in Euclidean space:
\begin{subequations}\label{gen-sep}
\be\label{totalE}
H = \frac{1}{2}\, (P_1^2+P_2^2)+h(Q_1,Q_2)
\ee
and require a quadratic (in momenta) first integral
\be\label{genF}
F = T_f+f(Q_1,Q_2) = (F_{11}(Q_1,Q_2) P_1^2+2F_{12}(Q_1,Q_2) P_1 P_2+ F_{22}(Q_1,Q_2) P_2^2) + f(Q_1,Q_2).
\ee
\end{subequations}
The formula $\{H,F\}=0$ has cubic and linear parts, which must separately vanish:
$$
\{T_e,T_f\}=0,\quad \{T_e,f\}+\{h,T_f\}=0.
$$
The first of these is equivalent to asking for the matrix $F_{ij}$ to define a second order Killing tensor of the metric corresponding to the kinetic energy $T_e$.  For a general kinetic energy, finding such a Killing tensor could be a difficult (and generally impossible) task.  However, it is well known that for a \underline{flat} metric, all second (and higher) order Killing tensors are built from tensor products of Killing vectors (see \cite{74-7}). In our context, this means that the quadratic part of $F$ is just some {\em quadratic form} of the functions $K_i$, given in (\ref{Euclid}), so there is a $6-$parameter family of them.  Two such quadratic forms are equivalent if they differ by a multiple of $T_e$.

The existence of a single quadratic integral corresponds to {\em separability} (in some coordinate system which diagonalises {\em both} $H$ and $F$).  In such a case, $h(Q_1,Q_2)$ (and the corresponding $f(Q_1,Q_2)$) depends upon two arbitrary functions (each of a single variable). The existence of \underline{two} quadratic integrals (the {\em superintegrable case}) restricts these arbitrary functions, so that the potential function is {\em fixed} up to a finite number of arbitrary {\em parameters}.

\subsection{Parabolic Coordinates}

The choice of separation variables is purely a property of the \underline{quadratic} part of $F$.  The corresponding pair of quadratic forms are  simultaneously diagonalised in these separation coordinates.   The examples in this paper are chosen in order to include those of \cite{15-1} as special cases.  They each have two quadratic integrals, one of which has the form
\be\label{k2k3}
F_1=T_1+f_1(Q_1,Q_2),\quad\mbox{where}\quad T_1=K_2 K_3=Q_2 P_1P_2-Q_1 P_2^2,
\ee
corresponding to (see \cite{90-16}) parabolic coordinates $(u,v)$, defined by
\begin{subequations}\label{Qu-tran}
\be\label{uv}
u = \frac{1}{2} (Q_1+\sqrt{Q_1^2+Q_2^2}),\;\;\; v = \frac{1}{2} (-Q_1+\sqrt{Q_1^2+Q_2^2}),\quad\mbox{with inverse} \quad
      Q_1 = u - v, \;\;\;  Q_2 = 2 \sqrt{u v}.
\ee
The corresponding canonical transformation is generated by
\be\label{SuQ}
S = (u - v) P_1 + 2 \sqrt{u v}\; P_2 \quad\Rightarrow\quad  Q_1 = u - v, \;\;  Q_2 = 2 \sqrt{u v},\;\;
      P_1 = \frac{u p_u - v p_v}{u + v}, \;\; P_2 = \frac{\sqrt{u v}\, (p_u + p_v)}{u + v},
\ee
\end{subequations}
which leads to
\begin{subequations}
\be\label{Kiuv}
K_1 = \frac{u p_u - v p_v}{u + v},\;\;\; K_2 = \frac{\sqrt{u v}\,(p_u + p_v)}{u + v},\;\;\; K_3 = \sqrt{u v}\, (p_u - p_v),
\ee
and
\be\label{TuT1}
T_h = \frac{1}{2}\, (K_1^2+K_2^2) = \frac{1}{2} \left(\frac{up_u^2 + vp_v^2}{u+v}\right),\quad T_1=K_2 K_3 = \frac{u v(p_u^2 - p_v^2)}{u + v},
\ee
\end{subequations}
which are evidently simultaneously diagonalised.  It is a simple matter to determine the separable potentials, satisfying $\{T_h+h(u,v),T_1+ f_1(u,v)\}=0$.  The results are well known (see \cite{90-16}):
\be\label{gensep}
H = \frac{1}{2} \left(\frac{up_u^2 + vp_v^2}{u+v} + \frac{A(u) + B(v)}{u + v}\right),\quad
        F_1 = \frac{u v(p_u^2 - p_v^2)}{u + v} + \frac{v A(u) - u B(v)}{u + v},
\ee
where $A(u),\, B(v)$ are arbitrary functions of their respective single variables.

\subsection{Superintegrable Restriction with $F_2 =  K_1 K_3 +f_2(u,v)$}

If we require that a second (independent) function Poisson commutes with $H$, then the arbitrary functions $A(u)$ and $B(v)$ are specified up to a finite number of parameters.  In the case
\begin{subequations}\label{super-1}
\be\label{F2=k1k3}
F_2 =  K_1 K_3 +f_2(u,v) = \frac{\sqrt{u v}\, (p_u - p_v)\,(u p_u - v p_v)}{u + v} +f_2(u,v),
\ee
the coefficients of $p_u$ and $p_v$, in $\{H,F_2\}=0$, give formulae for the first derivatives of $f_2(u,v)$, with integrability condition:
$$
2 u A''(u)+A'(u) = 2 v B''(v) + B'(v).
$$
This separable equation is easily solved, after which the (now consistent) overdetermined system for $f_2(u,v)$ can also be solved.  The result is
\bea
&& h(u,v) = \frac{1}{2} \left(\frac{2k_0+k_1\sqrt{u}+k_2\sqrt{v}}{u+v}\right),\quad
           f_1(u,v) = \frac{k_0(v-u) +k_1 v \sqrt{u}-k_2u \sqrt{v}}{u + v},\nn\\
           &&     \label{pots-1}  \\
&&    f_2(u,v) = \frac{2k_0 \sqrt{uv}+\frac{1}{2}(u-v)(k_1 \sqrt{v}-k_2 \sqrt{u})}{u + v}. \nn
\eea
\end{subequations}
With these potentials we have
\begin{subequations}\label{p-alg-1}
\be\label{f123}
\{H,F_1\}=0,\quad \{H,F_2\}=0,\quad\mbox{and define}\quad F_3=\{F_1,F_2\} ,
\ee
which is a cubic integral, which cannot be written as a linear (or polynomial) combination of $H,F_1$ and $F_2$ (but cannot, of course, be independent of them).  However,
\be\label{f1f3}
\{F_1,F_3\} = 2 F_2 H +\frac{1}{4} k_1 k_2,\quad \{F_2,F_3\} = -2 F_1 H +\frac{1}{8} (k_1^2- k_2^2),
\ee
forming a polynomial Poisson algebra.  As a consequence of these Poisson relations (ie not using the specific representation), we find the Casimir function:
\be\label{cas1}
{\cal C} = 2 H (F_1^2+F_2^2)- F_3^2 +\frac{1}{4} (k_2^2-k_1^2) F_1 +\frac{1}{2}k_1k_2 F_2.
\ee
\end{subequations}
Since the entire algebra is built as a symmetry algebra of $H$, this Casimir should be a function of $H$.  Indeed, using the specific representation, we find
$$
{\cal C} =  2 k_0^2 H+\frac{1}{4} k_0(k_1^2+k_2^2),
$$
which furnishes us with the expected algebraic relation between the four functions $H, F_i$.

\br[A second parabolic coordinate system]
Clearly, if we consider the pair $T_h$ and $T_2 = K_1K_3$, we obtain another parabolic coordinate system $(\bar u,\bar v)$, corresponding to the interchange $Q_1 \leftrightarrow Q_2$, and now
$$
T_h = \frac{1}{2}\, (K_1^2+K_2^2) = \frac{1}{2} \left(\frac{\bar u p_{\bar u}^2 + \bar v p_{\bar v}^2}{\bar u+\bar v}\right),
     \quad T_2=K_1 K_3 = \frac{\bar u \bar v(p_{\bar v}^2 - p_{\bar u}^2)}{\bar u + \bar v}.
$$
If we compose these transformations, we obtain
$$
\bar u = \frac{1}{2} (\sqrt{u}+\sqrt{v})^2,\quad \bar v = \frac{1}{2} (\sqrt{u}-\sqrt{v})^2,
$$
and the corresponding canonical transformation is a concrete realisation of the Lie algebraic automorphism
$$
\bar K_1 = K_2,\quad \bar K_2 = K_1,\quad \bar K_3 = -K_3 .
$$
As well as correctly transforming the \underline{quadratic} parts of $H, F_i$, this transformation also correctly transforms the potential functions, so
$$
(H,F_1,F_2)\mapsto (H,-F_2,-F_1)\quad\mbox{with}\quad (k_0,k_1,k_2)\mapsto \left(k_0,\frac{1}{\sqrt{2}}(k_1+k_2),\frac{1}{\sqrt{2}}(k_1-k_2)\right),
$$
which implies $F_3\mapsto -F_3$ and represents an automorphism of the above quadratic algebra.
\er

\subsection{Superintegrable Restriction with $F_2 =  K_1 K_2 +f_2(u,v)$}

In the case
\begin{subequations}\label{super-2}
\be\label{F2=k1k2}
F_2 =  K_1 K_2 +f_2(u,v) = \frac{\sqrt{u v}\, (p_u + p_v)\,(u p_u - v p_v)}{(u + v)^2} +f_2(u,v),
\ee
a similar calculation of $\{H,F_2\}=0$, with $H$ given by (\ref{gensep}), leads to
\be \label{pots-2}
 h(u,v) = \frac{1}{2} k (u-v),\quad  f_1(u,v) = k uv,\quad  f_2(u,v) = k \sqrt{u v},
\ee
\end{subequations}
from which we immediately see that $\{K_2,H\}=0$.  We then find
\begin{subequations}\label{p-alg-2}
\be\label{f1f2}
\{F_1,F_2\}= 2 K_2 (H-K_2^2),\quad \{K_2,F_1\}=-F_2,\quad  \{K_2,F_2\}=-\frac{k}{2}.
\ee
The functions $H, K_2, F_1, F_2$ cannot, of course, be independent.  They satisfy the algebraic relation
\be\label{cas1}
F_2^2 = 2 H K_2^2-K_2^4+k F_1
\ee
\end{subequations}

\section{Canonical Transformation to $(r,\varphi)$ Coordinates}\label{sect-polar}

We now write the results of Section \ref{sect-sep} in the original $(r,\varphi)$ coordinates and compare them with the systems derived in \cite{15-1}.

Combining the coordinate transformations (\ref{Qr-tran}) and (\ref{uv}), we obtain
\begin{subequations}\label{ur-tran}
\be\label{uvr}
u = \frac{2\, r^{1-n/2}}{2-n} \, \cos^2 \frac{(n-2)\varphi}{4},\quad
              v = \frac{2\, r^{1-n/2}}{2-n} \, \sin^2 \frac{(n-2)\varphi}{4},
\ee
with the corresponding canonical transformation generated by
\be\label{Sur}
S = \frac{2\, r^{1-n/2}}{2-n}  \, \left(p_u \cos^2 \frac{(n-2)\varphi}{4}+p_v \sin^2 \frac{(n-2)\varphi}{4}\right).
\ee
\end{subequations}
In these coordinates, $K_1=P_1$ and $K_2=P_2$ (defined by (\ref{p1p2})) and $K_3=\frac{2 p_\varphi}{(2-n)}$.
We can now write the superintegrable systems (\ref{super-1}) and (\ref{super-2}) in these coordinates.

The system (\ref{super-1}) takes the form
\bea
&&  H=\frac{1}{2}\, r^n \left(p_r^2+\frac{p_\varphi^2}{r^2}\right) + h(r,\varphi),\nn\\
&&  F_1= \frac{2}{2-n}\,\left(r^{n/2-1} p_\varphi \left(r p_r\, \sin \left(\frac{(n-2)\varphi}{2}\right)-
                   p_\varphi\, \cos \left(\frac{(n-2)\varphi}{2}\right)\right)+f_1(r,\varphi)\right), \nn\\
&&  F_2= \frac{2}{2-n}\,\left(r^{n/2-1} p_\varphi \left(r p_r\, \cos \left(\frac{(n-2)\varphi}{2}\right)+
                   p_\varphi\, \sin \left(\frac{(n-2)\varphi}{2}\right)\right)+f_2(r,\varphi)\right),\nn
\eea
where
\bea
h(r,\varphi) &=& c_1 r^{(n-2)/2} + r^{(n-2)/4} \left(c_2 \cos \frac{(n-2)\varphi}{4}
                       + c_3 \sin \frac{(n-2)\varphi}{4}\right),\nn\\
f_1(r,\varphi) &=& -c_1 \cos \frac{(n-2)\varphi}{2}-r^{(2-n)/4}\, \sin \frac{(n-2)\varphi}{2}
    \left(c_3 \cos \frac{(n-2)\varphi}{4}  -c_2 \sin \frac{(n-2)\varphi}{4}\right),  \nn\\
f_2(r,\varphi) &=& c_1 \sin \frac{(n-2)\varphi}{2}+ \frac{1}{2}\, r^{(2-n)/4} \,\cos \frac{(n-2)\varphi}{2}
                             \left(2 c_2  \sin \frac{(n-2)\varphi}{4}  -2c_3 \, \cos \frac{(n-2)\varphi}{4}\right).  \nn
\eea

\br
Cases 1 and 2 of \cite{15-1} (see their list (3.8)) correspond to $c_1=c_3=0$, with $n=6$ and $n=-2$, respectively.
\er

The system (\ref{super-2}) takes the form
\bea
&&  H=\frac{1}{2}\, r^n \left(p_r^2+\frac{p_\varphi^2}{r^2}\right) + h(r,\varphi),\nn\\
&&  F_1= \frac{2}{2-n}\,\left(r^{n/2-1} p_\varphi \left(r p_r\, \sin \frac{(n-2)\varphi}{2}-
                   p_\varphi\, \cos \frac{(n-2)\varphi}{2}\right)+f_1(r,\varphi)\right), \nn\\
&&  F_2= \frac{1}{2}\, r^{n}  \left(\left(p_r^2-\frac{p_\varphi^2}{r^2}\right)\,\sin (n-2)\varphi
                       -\frac{p_r p_\varphi}{r}\, \cos (n-2)\varphi\right)  +f_2(r,\varphi),\nn
\eea
where
$$
h= c_1 r^{1-n/2}\, \cos\frac{(n-2)\varphi}{2}, \quad f_1=c_1 r^{1-n/2}\, \sin\frac{(n-2)\varphi}{2}, \quad
   f_2= -\frac{1}{2} c_1 r^{2-n}\, \sin^2\frac{(n-2)\varphi}{2}.
$$

\br
Cases 3 and 4 of \cite{15-1} (see their list (3.8)) correspond to $n=0$ and $n=4$, respectively.
\er

\br
The selection of the $4$ values of $n$ is just a consequence of the restriction to the case $U(\varphi)=-\cos\varphi$, in \cite{15-1}.  Comparing with the general formulae above, this just means choosing $n$ so that $(n-2)/4=\pm 1$ or $(n-2)/2=\pm 1$.
\er

%\bibliography{apf}

\end{document}